# Ultrafast transient absorption spectroscopy of 2D semiconductors: a review

Yuri D. Glinka[1,2, a)]

[1] *The institute of Optics, University of Rochester, Rochester, NY 14627, USA*
[2] *Institute of Physics, National Academy of Sciences of Ukraine, Kyiv 03028, Ukraine*

[a)]Electronic mail: yuridglinka@yahoo.com

**Abstract**
Despite the decades that have passed since the discovery of ultrafast transient absorption spectroscopy and its apparent simplicity, this method is still often subject to experimental errors and misinterpretations when applied to 2D semiconductors. The reason for this lies not only in the unique nature of these extremely thin samples, but also in the different experimental configurations and data processing methods used. Moreover, since this type of spectroscopy was originally used to characterize the ultrafast relaxation dynamics of photoexcited carriers in chemical compounds, colloidal nanostructures and gas molecules, a purely "molecular" approach to interpreting transient absorption spectra of 2D semiconductors is often proposed. However, this approach is fundamentally inapplicable to thin-film semiconductors grown or mechanically exfoliated on transparent substrates. This review considers the recent experimental results of ultrafast transient absorption spectroscopy of 2D semiconductors in a wide spectral range from several THz to UV (~1000 THz) based on the "solid-state" approach to their interpretation. We also highlight typical errors that arise in measuring, processing and interpreting transient absorption spectra of 2D semiconductors.

**Introduction**

Ultrafast transient absorption spectroscopy is a type of pump-probe technique made possible by several major advances in ultrafast laser physics and nonlinear optics [1, 2]. These include the generation of femtosecond optical pulses in mode-locked lasers, as well as the generation of broadband terahertz (THz) and mid-IR - UV supercontinuum pulses in a nonlinear optical medium. Consequently, femtosecond resolution can be achieved for tracking a broadband pulse that probe a sample after it is excited by a narrowband pump pulse. This experimental approach thus allows simultaneous time-resolved measurements of pump-induced changes in the absorption spectrum of a material over a wide range of wavelengths [3, 4]. Accordingly, this method was applied to study the ultrafast dynamics of photoexcited carriers in thin-film semiconductors in the range from several THz to UV [5-10].

There are two fundamentally different approaches to the theoretical description of these dynamics. The first approach considers first the physical phenomena caused by injected free carriers in bulk semiconductors and then the influence of these phenomena on their optical properties, determined by the complex refractive index, $\tilde{n} = n + i\kappa$, where $n$ is the refractive index and $\kappa$ is the extinction coefficient, related to the absorption coefficient as $\alpha = 4\pi\kappa/\lambda$, where $\lambda$ is the wavelength [7, 8]. This theory was developed for free carriers regardless of whether they were injected into bulk semiconductors electrically or optically, and considers three physical phenomena: bandfilling, bandgap shrinkage (renormalization), and free carrier absorption [11].

The second approach first considers the ultrafast transient absorption spectra of thin-film semiconductors within the framework of the Drude-Lorentz model and then tries to explain them using suitable physical phenomena [9, 10]. Moreover, in this case, since transient absorption spectroscopy was originally used to characterize the ultrafast relaxation dynamics of photoexcited carriers in chemical compounds, colloidal nanostructures and gas molecules, a purely "molecular" interpretation of the experimental results is usually applied. Physical phenomena that are typically considered within the "molecular" approach include ground-state bleaching, stimulated emission, excited-state absorption, and product absorption [3, 4]. Thus, the two approaches mentioned concern completely different physical phenomena, although they are often used to interpret the same experimental results.

Due to limitations in light penetration into semiconductors, ultrafast transient absorption spectroscopy can only be applied to very thin semiconductor films to minimize unwanted distortion, saturation and reabsorption effects. From this point of view, 2D semiconductors are ideally suited to this criterion, since the penetration depth of light into bulk analogs of 2D semiconductors significantly exceeds their thickness [12, 13]. In general, 2D semiconductors often consist of layers of atoms held together by strong covalent bonds within each layer, with the layers remaining free of dangling bonds and held together by much weaker van der Waals forces. Examples include graphene (a single layer of carbon atoms) [14-19], monolayers of transition metal dichalcogenides (TMDCs, e.g. $MoS_2$, $WS_2$) [20-35], boron nitride (h-BN) [36], phosphorene [37], topological insulators (TI, e.g. $Bi_2Se_3$, $Sb_2Te_3$) [38-53] and other 2D semiconductor structures [8, 54, 55].

2D semiconductors have unique properties such as potentially high carrier mobility and thermal conductivity, as well as optical transparency and the ability to tune the bandgap by strain, external electric and magnetic fields, and stacking multiple layers [31, 32]. Although structural defects strongly influence the transport properties of 2D semiconductors [56], by tuning the fundamental physical characteristics of high-quality monolayers, even higher carrier mobilities can be achieved than in bulk silicon [57]. These physical characteristics include a low effective mass, a high speed of sound, a high optical phonon frequency, a low Born charge-to-polarizability ratio, and a large carrier-to-lattice distance. Consequently, 2D semiconductors show great potential for creating next-generation electronic and optoelectronic devices that can be used in a wide range of applications, including transistors, sensors, energy storage and quantum technologies.

It is important to distinguish between 2D semiconductors and semiconductor quantum wells (QWs). Both semiconductor nanostructures exhibit quantum confinement of carriers, which suggests the discrete nature of the conduction/valence band, consisting of a series of the energy subbands [58–61]. However, within each subband, electrons/holes can still move and have a range of kinetic energies, resulting in a continuum of available states with respect to the in-plane carrier momentum [62]. Since this continuous range can overlap with the corresponding ranges of higher energy subbands, a quasi-continuous conduction/valence band with a joint density of states is formed. Accordingly, the conduction and valence bands of both semiconductor nanostructures, as well as the exciton states, are usually represented as parabolas, indicating that the carrier energy varies as a function of the square of their in-plane momentum (effective mass approximation) (Fig. 1) [62]. However, both bands may have a more global internal structure. In semiconductor QWs, the hole subband energies are further split by strain [62], whereas in 2D semiconductors such as TMDCs, the splitting is determined by strong spin-orbit coupling and inversion symmetry breaking [20-24]. The electron subband energies can also be split due to the Rashba effect [8]. The quasi-continuous nature is also characteristic of linearly dispersive Dirac surface states in TIs [38-53], for which extremely high in-plane mobility of massless Dirac fermions was observed [63].

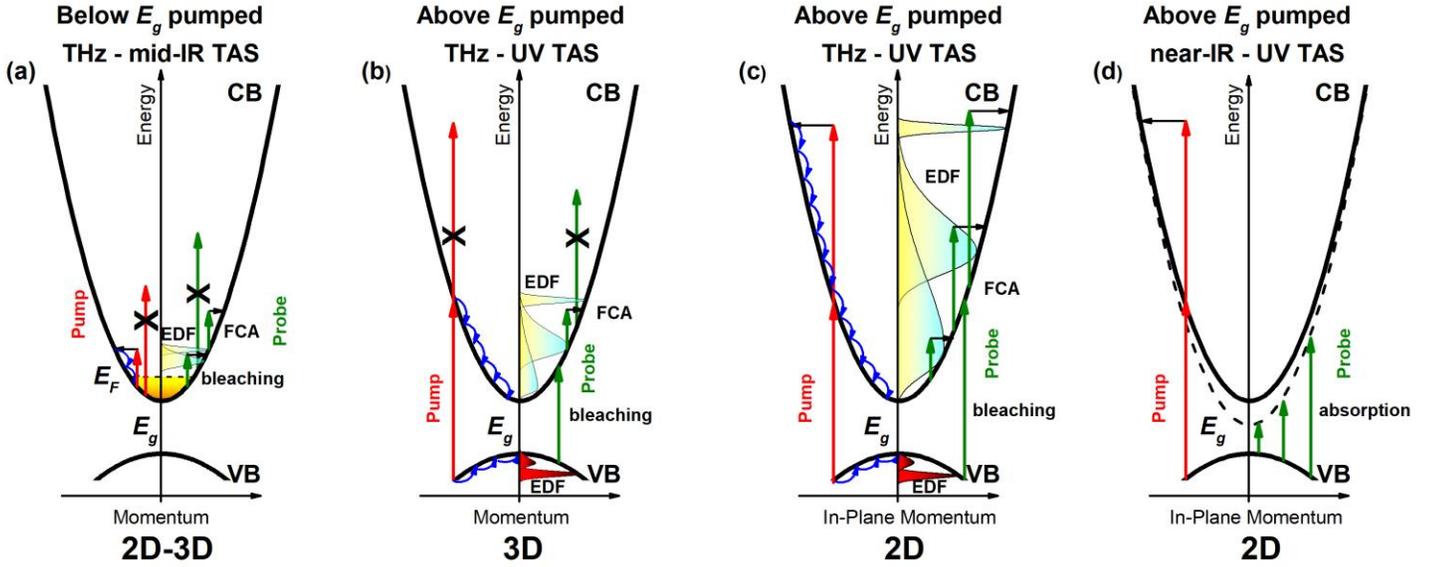

FIG. 1. Schematic diagram of a semiconductor with bandgap $E_g$ and different types of transient absorption spectroscopy (TAS) with pumping below $E_g$ and above $E_g$ for bulk (3D) and 2D semiconductors as indicated. The corresponding pump and probe transitions are shown with red and green arrows, respectively. The cascade emission of phonons by hot electrons and holes in the conduction and valence bands (CB and VB, respectively) is shown schematically, along with the typical evolution of the electron/hole energy distribution function (EDF). (a) TAS in the THz - mid-IR range with pumping below $E_g$; only low-energy transitions due to Drude-type free carrier absorption (FCA) and Drude-type FCA bleaching are allowed for both 2D and 3D semiconductors to conserve the electron momentum. (b) TAS in the THz - UV range for 3D semiconductors pumped above $E_g$; in the THz - mid-IR range only Drude-type FCA contributes, but in the near-IR - UV range the main contribution is due to the CB absorption bleaching at energies below the pump photon energy. The momentum-related restrictions of multiphoton (e.g., two-photon) pumping and high-energy Drude-type FCA transitions are shown. (c) TAS in the THz - UV ranges for 2D semiconductors pumped above $E_g$ by multiphoton excitation; in the THz - Vis range Drude-type FCA and inverse-bremsstrahlung-type FCA contribute, but in the near-IR - UV range the main contribution is due to the CB absorption bleaching at energies below the multiphoton pump energy. (d) TAS in the near-IR - UV range for 2D semiconductors pumped above $E_g$ by multiphoton excitation. The bandgap renormalization effect results in absorption over a wide spectral range.

One of the important characteristics of electronic and optoelectronic devices is the energy distribution function (EDF) of carriers within the conduction/valence band on a subpicosecond time scale. To estimate EDF, a carrier energy-sensitive method such as time- and angle-resolved photoemission spectroscopy (TrARPES) is usually used [16–19, 26, 29, 38–42]. In contrast, although transient absorption spectroscopy has been widely used to study ultrafast carrier dynamics in 2D semiconductors [15, 20–25, 27, 28, 30, 43–55], the evolution of transient absorption spectra on the subpicosecond time scale has not been sufficiently addressed. However, it has recently been shown that such changes in transient absorption spectra can also be used to estimate the intraband EDF of carriers through the dynamics of their population in the conduction or valence band [45–50].

Due to the different nature of the interaction of the probe light with pump-excited 2D semiconductors in different spectral regions, the corresponding changes in the absorption spectrum can be positive (pump-induced absorption) or negative (pump-induced transmission or absorption bleaching). Within the "solid-state" approach considering the effects caused by injected free carriers [11], the positive contribution can be due to two reasons: (i) photoexcited carriers absorb light under their collective excitation, a process known as Drude-type free carrier absorption (FCA) [62]; (ii) the pump-excited carrier population as a whole induces new unoccupied states in the conduction band due to its shift toward lower energies, a process known as bandgap renormalization [7, 8]. In addition, another positive contribution has been proposed for 2D semiconductors. Since the mean free path of photoexcited carriers is much larger than the thickness of 2D semiconductors [64–67], pump-excited carriers can also absorb light when colliding with potential barriers at the boundaries of the 2D semiconductor, a process known as inverse-bremsstrahlung-type FCA [8, 45–50].

In contrast, there is only one negative contribution (absorption bleaching, also known as Burstein-Moss shift or Pauli blocking), which is based on Fermi-Dirac statistics and reflects the population dynamics of pump-excited carriers. Accordingly, these carriers fill the conduction/valence band and, therefore, transiently block the electron transitions that could be initiated by the probe light (Pauli blocking) [7, 8, 20–25, 27, 28, 30, 43–55].

This "solid-state" classification of changes occurring in the absorption spectrum of 2D semiconductors upon pumping is principally different from the "molecular" classification mentioned above. Moreover, the appearance of different contributions to the transient absorption spectrum of 2D semiconductors strongly depends on whether the pump and probe photon energy is less or greater than the bandgap energy ($E_g$). Accordingly, intraband or interband optical transitions can be involved [62], thus determining contributions of different nature to the transient absorption spectra (Fig. 1) [50]. Despite the fundamental difference between the "solid-state" and "molecular" approaches, both have been used to characterize the transient absorption spectra of thin-film semiconductors [9, 10, 50, 51, 68]. Mixing these approaches often introduces confusion not only into terminology but also into the interpretation of experimental results. For example, when using the "molecular" approach to characterize transient absorption spectra caused by exciton dynamics, the term "ground-state bleaching" sounds absurd, since the exciton is a quasiparticle that exists only in an excited state.

Another source of confusion relates to the modulation technique commonly used. Since the pump-induced changes in the absorption spectrum are usually very small, ultrafast transient absorption spectroscopy setups typically include a chopper to modulate the pump beam and a lock-in amplifier operating at the chopper frequency to improve the signal-to-noise ratio. Accordingly, the measurement philosophy is to analyze the logarithm of the ratio of two intensities $\ln(I_{\nu 0}/I_\nu)$ (optical density) of a THz broadband or mid-IR - UV supercontinuum radiation transmitted through a 2D semiconductor at a certain frequency ($\nu$) without and with a pump pulse. This experimental arrangement means that the sign of the transient absorption spectrum can be easily changed by adjusting the chopper modulation phase. Accordingly, the actual sign of the transient absorption spectrum is ambiguous and must be chosen considering the nature of the corresponding physical processes and using a calibration sample with well-known transient absorption properties [50].

It should also be noted that instead of transient absorption spectroscopy, transient reflectivity spectroscopy followed by the Kramers-Kronig transform is often used to characterize the transient absorption properties of 2D semiconductors [9, 10, 25, 51]. However, in this case, the shape and intensity of the measured spectrum depend significantly on the angle of incidence of the probing light and represent a composition of its multiple reflections from the boundaries of the sample [50]. Moreover, depending on whether the Kramers-Kronig transformation is applied, the sign of the contributions can be represented as coinciding or opposite to the sign of the contributions in the directly measured transient absorption spectra [25, 28, 30, 51-53]. Although the Kramers-Kronig relations do not change the sign of the optical response [69], this discrepancy creates additional confusion in the interpretation of experimental results.

Furthermore, transient absorption spectra measured in the subpicosecond range are usually convolved with the chirp of the supercontinuum probe pulse [3, 48, 50]. The chirp effect, which results in a frequency shift over the pulse duration, typically causes the zero-time shifting to longer times with increasing probe wavelength. To eliminate this effect, a typical post-experimental procedure is to numerically remove any zero-time shift with the probe pulse wavelength. However, the transient absorption spectrum on this time scale also reveals ultrafast carrier cooling dynamics due to the emission of a phonon cascade [8, 13, 39, 45, 65, 67]. Accordingly, the chirp extraction procedure becomes extremely important for obtaining the real relaxation dynamics of photoexcited carriers on a subpicosecond time scale.

In this review, we consider recent experimental results of ultrafast transient absorption spectroscopy of 2D semiconductors in a wide spectral range from several THz to UV (~1000 THz), using a "solid-state" approach to their interpretation. We first review the general relaxation trends of photoexcited carriers in 2D semiconductors and then discuss the contributions of specific physical phenomena to transient absorption spectra. Finally, we highlight the most common errors that arise when measuring, processing and interpreting ultrafast transient absorption spectra of 2D semiconductors.

## 2. General tendencies of relaxation of photoexcited carriers in 2D semiconductors.

Since the physical phenomena that determine the contribution to the transient absorption spectrum are different in different spectral regions, the corresponding experimental configurations are required to study them. Accordingly, transient absorption spectroscopy of 2D semiconductors can be divided into three directions depending on the experimental configurations used: (i) below-bandgap pumping and below-bandgap probing [Fig. 1(a)]; (ii) above-bandgap pumping and below-bandgap probing [Fig. 1(b) and (c)]; (iii) above-bandgap pumping and above-bandgap probing [Fig. 1(b), (c) and (d)]. To discuss the physical phenomena more specifically, we first consider some of the most common examples for each of the experimental configurations illustrating the general relaxation trends of photoexcited carriers in 2D semiconductors.

*2.1 Below-bandgap pumping and below-bandgap probing.*

We start with the simplest case, when pumping and probing are below the bandgap. This experimental configuration can only be applied to doped 2D semiconductors (e.g. *n*-type). In this case, as in metals, light absorption occurs due to intraband optical transitions when extrinsic electrons in the conduction band are excited from filled states below the Fermi energy ($E_F$) to empty states above it [Fig. 1(a)] [62]. The absorption of light by pump-excited extrinsic electrons typically occurs in the THz - mid-IR range and is commonly referred to as Drude-type FCA. The pumping and probing range for a given experimental configuration is strictly limited by the plasma frequency (or electron density, $n_e$) and the electron scattering rate. The latter determines the oscillator strength of intraband electronic transitions, since it guarantees the conservation of electron momentum during optical transitions [62]. Further thermalization of pump-excited electrons at a certain electron temperature ($T_e$) takes several tens of femtoseconds and leads to their Fermi-Dirac distribution. The resulting contribution to the transient absorption spectrum is positive and can be observed as an initial increase in the transient conductivity signal [Fig. 2(a) and (f)] [5].

Since the pump-excited extrinsic electrons have high kinetic energy, their cooling occurs due to the emission of a phonon cascade [62]. The cooling process results in a redistribution of electrons towards lower energy states in the conduction band. This behavior manifests itself as a decay of the THz conductivity signal, which typically takes several picoseconds.

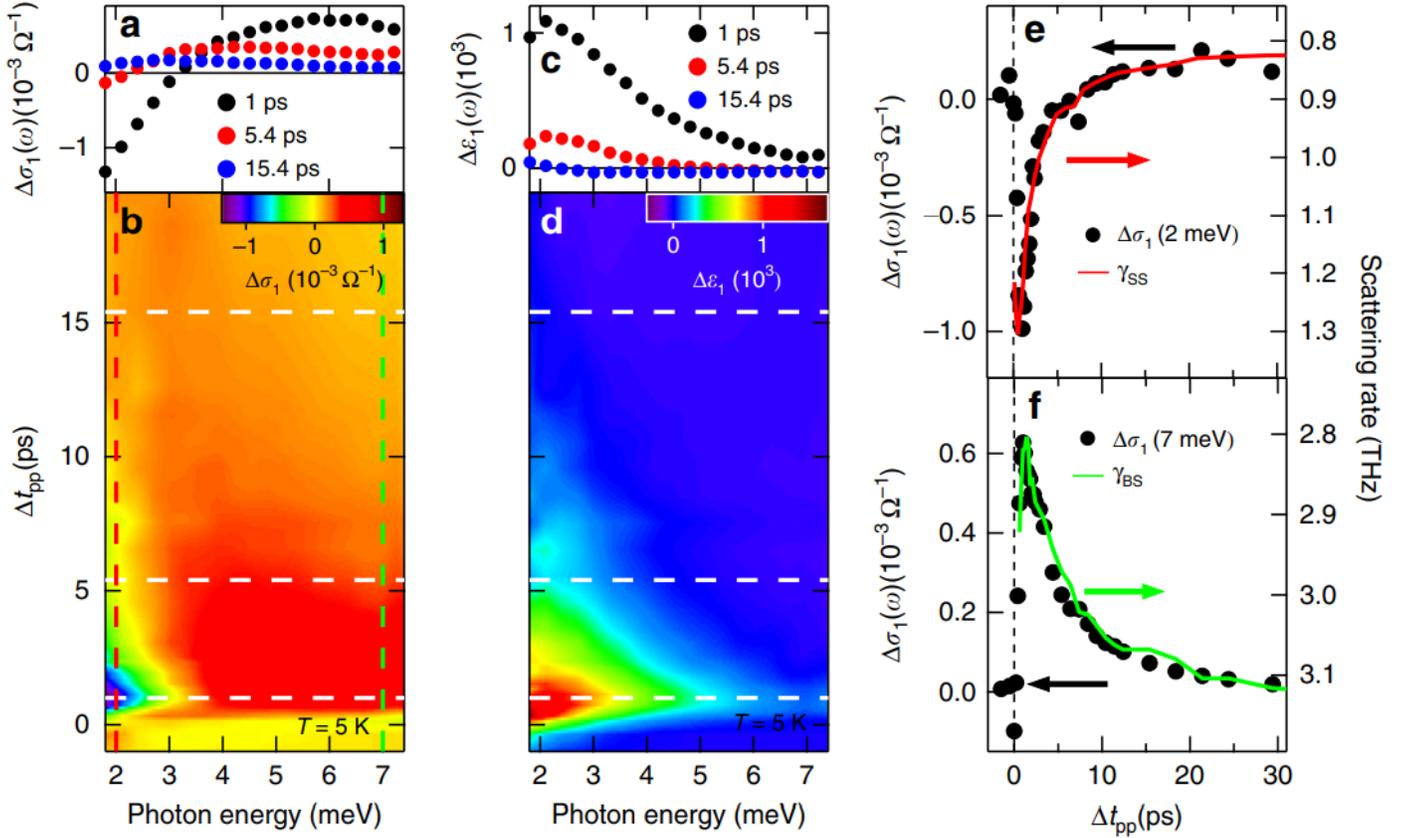

FIG. 2. Frequency-dependent dynamics of THz spectra at low temperature for a $Bi_2Se_3$ film. Pump-induced THz spectra of **a**, **b** $\Delta\sigma_1(\omega)$ (THz conductivity) and **c**, **d** $\Delta\varepsilon_1(\omega)$ (dielectric function) after 248 meV (5 μm) photoexcitation with fluence 12 μJ cm$^{-2}$ at T = 5 K as a function of pump-probe delay $\Delta t_{pp}$. **a** and **c** show the THz spectra from three cut positions from the corresponding 2D plots of **b** and **d**, respectively, at $\Delta t_{pp}$ = 1, 5.4, and 15.4 ps, as indicated by the white-dashed lines. **e** THz conductivity at 2 meV (black dots, left axis), as a function of $\Delta t_{pp}$, from the frequency-cut position in **b** as indicated by the red-dashed line. The scattering rate of the surface state $\gamma_{SS}$ (red curve, right axis) obtained from theoretical fitting is plotted together to compare relaxation dynamics. **f** Similar to **e**, THz conductivity at 7 meV (black dots, left axis) and the scattering rate of bulk state $\gamma_{BS}$ (green curve, right axis) are plotted. $\Delta\sigma_1$(7 meV) is from the frequency-cut position in **b** as indicated by the green-dashed line. The relaxation dynamics of the surface (bulk) scattering rate matches with that of THz conductivity at 2 meV (7 meV) very well. This frequency-dependent THz relaxation dynamics indicates surface (bulk) state is more sensitive to low (high) THz frequency. Reproduced from [5]. CC BY 4.0.

Empty states below $E_F$ can also contribute to the transient absorption spectrum, which is associated with Drude-type FCA bleaching [5]. The latter process is based on Fermi-Dirac statistics and the Pauli exclusion principle and appears as a negative contribution (negative THz conductivity) [Fig. 2(a) and (e)].

*2.2 Above-bandgap pumping and below-bandgap probing.*

This experimental configuration can be applied to intrinsic 2D semiconductors where carrier pumping is due to interband optical transitions [Fig. 1(b), (c) and (d)] [60]. In addition to the conventional one-photon pumping in 3D semiconductors [Fig. 1(b)], for 2D semiconductors the pumping process can be multiphoton (multistep), thus extending the transient absorption spectra towards higher energies [Fig. 1(c) and (d)] [8, 45–47, 49, 50]. This pumping regime is caused by the inverse-bremsstrahlung-type FCA, which arises due to collisions of photoexcited electrons with potential barriers at the boundaries of 2D semiconductors. The corresponding FCA process is maximal at the peak intensity of the pump pulse and is characterized by an extremely high carrier scattering rate [50]. Since this rate significantly exceeds all known carrier scattering rates for carrier thermalization, Auger heating and Auger recombination, carrier multiplication, carrier cooling due to phonon emission and electron-hole recombination [45, 70–74], multiphoton (multistep) pumping precedes all these carrier relaxation processes and does not affect their temporal dynamics.

The multiphoton (multistep) excitation approach based on the inverse-bremsstrahlung-type FCA [45, 48–50] contrasts with the Auger-type approach proposed for resonantly excited upconversion quantum emission [photoluminescence (PL)] from 2D TMDCs [31, 33–35]. The fundamental difference between these two processes is that the inverse-bremsstrahlung-type FCA is a non-resonant intraband process, whereas Auger pumping is a resonant interband process involving a specific three-carrier interaction. In the latter case, spectrally narrow PL emission was observed from quantum dots associated with specific locations of the flakes near the edges, folds, or nanobubbles in the monolayer. Although the detailed microscopic origin of these atom-like emitters in 2D TMDCs is still unclear, highly localized strain gradients are often correlated with the physical origin of these quantum emitters [31].

This behavior is in good agreement with the general approach developed for quantum dots, according to which even for 1D semiconductors (nanorods), the efficiency of Auger-type effects drops significantly and strongly depends on the actual carrier density in nanorods compared to that in quantum dots [70, 72]. In addition, due to the integrating effect over the entire laser spot, typically in the range of 1.0 to 100 μm, transient absorption spectroscopy is not as sensitive to local defects in 2D semiconductors as the quantum emission. Consequently, for transient absorption spectroscopy applied to flat and structurally perfect TMDC monolayers, all Auger-type resonant pumping mechanisms that are usually considered for the excitation of spatially localized PL and its upconversion can be ignored. Thus, multiphoton (multistep) pumping based on the inverse-bremsstrahlung-type FCA appears to be a more reliable mechanism for characterizing the transient absorption spectra of high-quality 2D semiconductors compared to the Auger-type upconversion mechanism developed for semiconductor quantum dots.

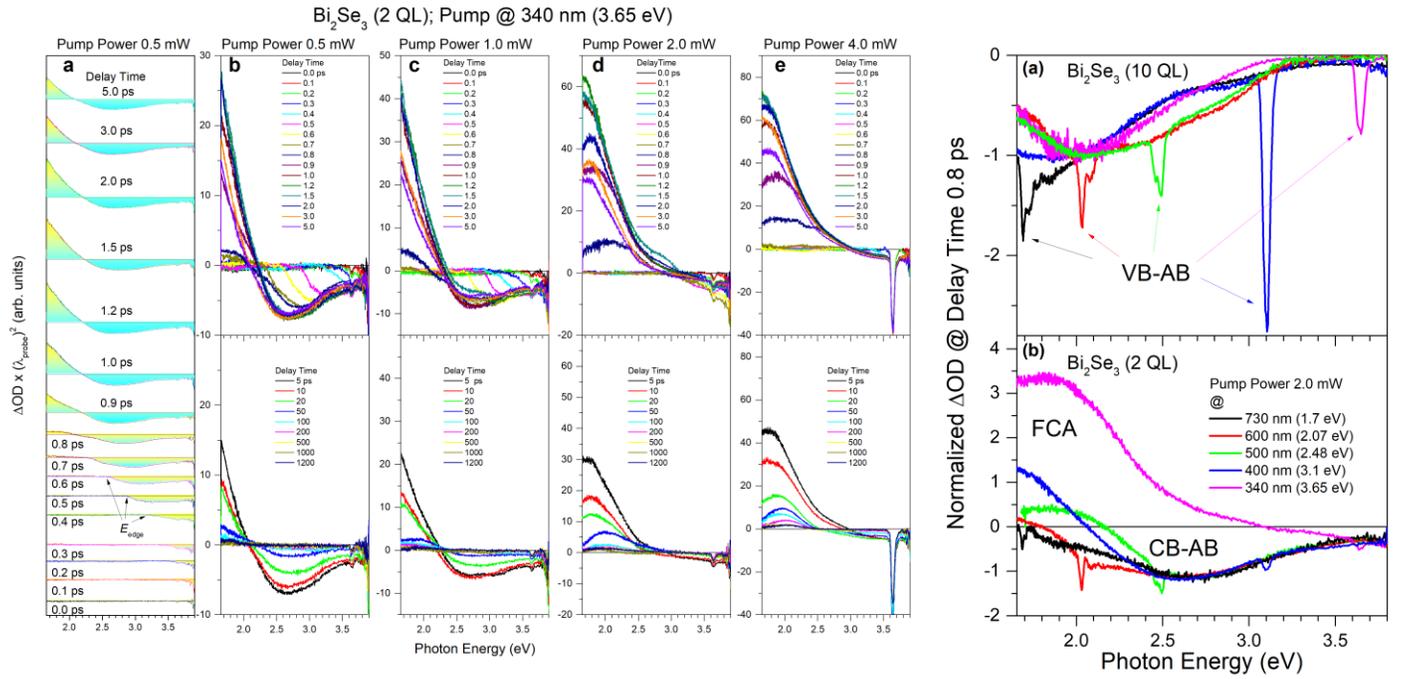

FIG. 3. (Left) TA spectra of the topologically trivial insulator phase of the 2D TI Bi$_2$Se$_3$. (**a** – **e**) Set of TA spectra of the 2 nm thick Bi$_2$Se$_3$ film measured at delay times indicated by the corresponding colors using the 340 nm pumping (∼ 3.65 eV photon energy) of different powers, as indicated for each of the columns. Part (**a**) and (**b**) show the same TA spectra for clarity. The zero-intensity lines of TA spectra in part (**a**) were shifted along the ΔOD axis for better observation. The factor ($\lambda_{probe}$)$^2$ in the ΔOD axis arises due to the transformation of wavelength-to-energy units. The low-energy edge of the transiently excited electron population ($E_{edge}$) is indicated in part (**a**). Reproduced from [48]. CC BY 4.0. (Right) (**a**) and (**b**) TA spectra of the 3D TI Bi$_2$Se$_3$ (10 nm thick) and the 2D TI Bi$_2$Se$_3$ (2 nm thick), respectively, measured at delay time of ∼0.8 ps using pump pulses of certain wavelengths, as indicated by the corresponding colors. The pump power for all wavelengths was ∼2 mW. Broadband CB absorption bleaching (CB-AB) and FCA contributions are shown along with narrow peaks associated with the VB absorption bleaching (VB-AB) contribution. Reproduced from [49]. © IOP Publishing Ltd. All rights reserved.

Transient absorption spectra measured for intrinsic 2D semiconductors using above-bandgap pumping in the near-IR - UV spectral range and below-bandgap probing in the THz - mid-IR spectral range are very similar to the spectra measured for doped 2D semiconductors using below-bandgap pumping and probing [6, 7]. The reason is that in both cases the spectra are caused by Drude-type FCA. In contrast, Drude-type FCA bleaching is not observed when pumping above the bandgap.

However, the contribution of the Drude-type FCA in intrinsic 2D semiconductors can be extended towards higher energies due to the inverse-bremsstrahlung-type FCA. As demonstrated for 2D TI Bi$_2$Se$_3$ ($E_g$ ∼ 0.3 eV) (Fig. 3) [45-49], the inverse-bremsstrahlung-type FCA contribution gradually increases and extends even to the visible range when both the pump photon energy and the photoexcited carrier density increase. Accordingly, the FCA range can sometimes exceed the bandgap energy of a 2D semiconductor.

The rise-time constant of the inverse-bremsstrahlung-type FCA signal is ∼0.3 ps. However, this time characterizes the rate of accumulation of carriers in Dirac surface states rather than the rate of their light absorption [45]. Moreover, this type of FCA manifests itself in transient absorption spectra with a characteristic delay of ∼1 ps after excitation of carriers, since their accumulation in Dirac surface states is required. This delay is completely determined by the initial energy of the carriers under pumping, their cooling rate, and the magnitude of the gap opening in the Dirac surface states with decreasing thickness of the 2D TI Bi$_2$Se$_3$. The initial decrease in the intensity of this contribution occurs with a decay-time constant of ∼4.0 ps, which is longer than the typical initial relaxation time of carriers in bulk states (∼2.0 ps) probed using the absorption bleaching contribution [45-49]. The reason for this is that different electron-phonon scattering mechanisms are involved, namely, quasi-elastic and inelastic electron-phonon scattering in Dirac surface states and bulk states, respectively. In contrast, the longer timescale dynamics due to electron-hole recombination are quite similar for both contributions (∼50 ps), although they become shorter with decreasing thickness of the 2D TI Bi$_2$Se$_3$.

*2.3 Above-bandgap pumping and above-bandgap probing.*

This experimental configuration is usually used in the near-IR - UV spectral range and is characterized by multiple positive and negative peaks in the transient absorption spectra (Fig. 4) [7-10, 20-25, 27, 28, 30, 45-49, 51-53, 68]. The multi-peak structure indicates complex carrier dynamics and often leads to the most controversial interpretations. When pumped above the bandgap, the initial carrier population exists as an electron-hole plasma. As we discussed in the previous section, the plasma contribution to the transient absorption spectrum can be monitored directly using the inverse-bremsstrahlung-type FCA. This contribution reflects the fastest optical response of 2D semiconductors, since other contributions are not yet activated. The reason for this is that the electron-hole plasma is electrically neutral and therefore does not change the complex refractive index of 2D semiconductors, and hence their optical properties. However, further thermalization of the electron-hole plasma, cooling of carriers and their recombination can be monitored with greater accuracy using other contributions. They include one positive contribution (bandgap renormalization) and one negative contribution (absorption bleaching) [Fig. 1(c) and (d)] [50]. Thus, these contributions can overlap with the positive contribution due to the inverse-bremsstrahlung-type FCA when higher pump powers are applied.

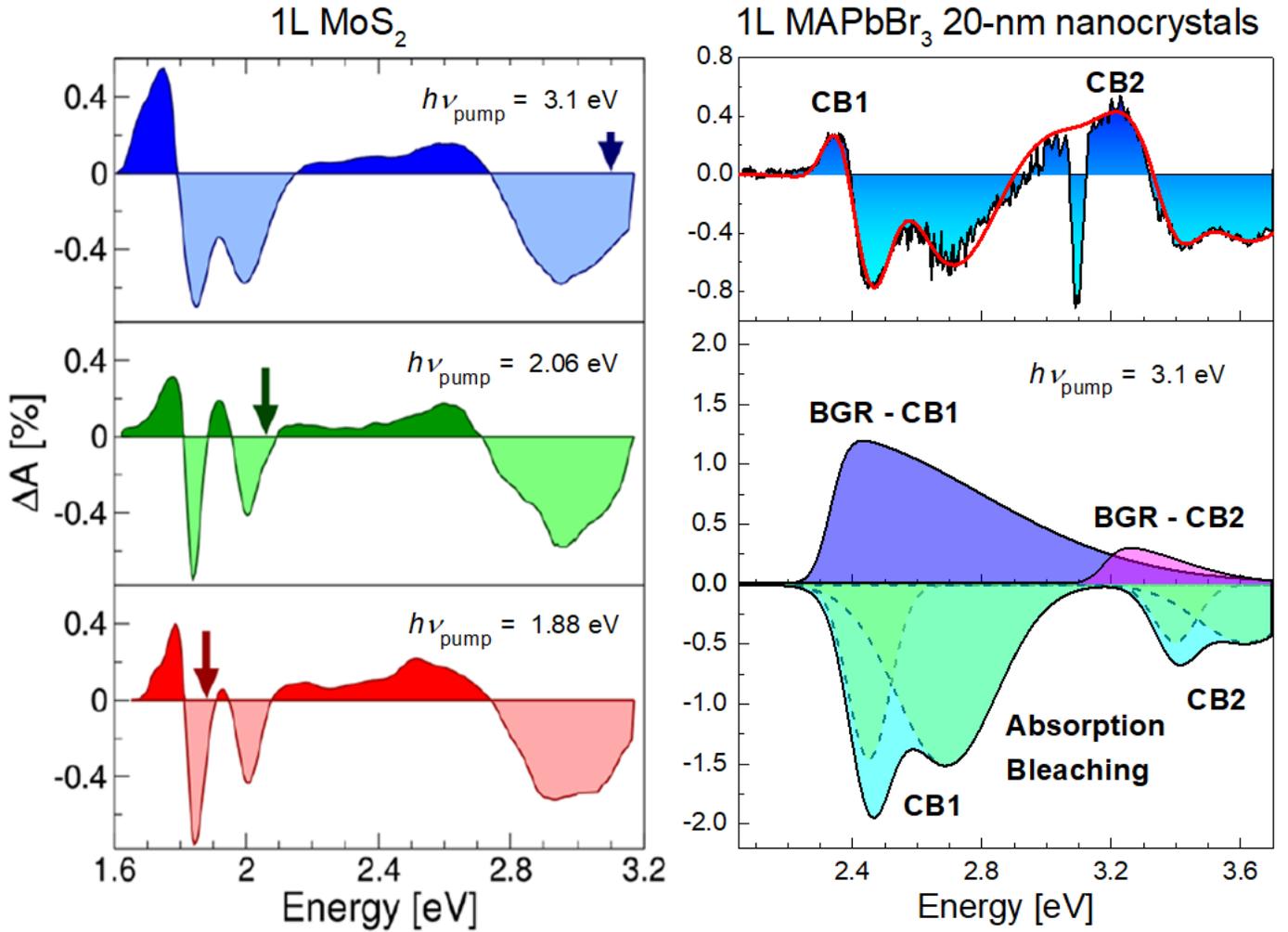

FIG. 4. (Left) Transient absorption spectra of 1L-MoS$_2$, recorded with a fixed pump-probe delay of 300 fs, for three pump photon energies, *i.e.* in resonance with the A ($h\nu_{pump}$ = 1.88 eV) and B ($h\nu_{pump}$ = 2.06 eV) excitons, and out of resonance with C ($h\nu_{pump}$ = 3.1 eV). Reprinted with permission from [21]. Copyright (2016) American Chemical Society. (Right) The transient absorption spectrum of 1L MAPbBr$_3$ 20-nm nanocrystals measured at a pump-probe delay of 700 fs using $h\nu_{pump}$ = 3.1 eV. The red thick curve shows the best fit to the data using two asymmetric Gaussians for the bandgap renormalization (BGR) contributions and four normal Gaussians for the absorption bleaching contributions. The corresponding contributions and fitting components are shown below for the two conduction subbands (CB1 and CB2). Reproduced from [8]. CC BY 4.0.

After the electron-hole plasma is thermalized due to carrier-carrier interactions, the carriers acquire their electron ($T_e$) and hole ($T_h$) temperature, which determines the corresponding Fermi-Dirac distributions. The subsequent relaxation dynamics and the resulting contribution to the transient absorption spectrum are determined solely by whether and how quickly excitons are formed. Specifically, due to the large asymmetry between the mobilities of electrons and holes, the thermalization of electron-hole plasma initially leads to dynamic charge separation (photo-Dember effect) [75]. The resulting transient electric field changes the complex refractive index of 2D semiconductors and hence contributes to their absorption spectrum. However, due to the decrease in dielectric screening and, consequently, the increase in the Coulomb interaction between electrons and holes in 2D semiconductors [58–61], the photo-Dember effect is significantly limited by the formation of excitons.

Accordingly, transient absorption spectra in the near-IR – UV spectral range can be considered within the framework of two models, namely, those associated with the transient electric field for free carriers or the Drude-Lorentz model (Lorentz oscillator model) for excitons [58, 61, 76]. In addition, the model based on Fermi-Dirac statistics can also be used to characterize the dynamics of carriers in 2D semiconductors [7, 8, 11, 45-50]. Within the framework of the latter model, probing of the states of quasi-continuous valence and conduction bands depends significantly on their filling with carriers excited by pumping. The corresponding dynamics completely determines the shape and intensity of the transient absorption spectrum via the Pauli blocking mechanism. This model explains the reason why the absorption bleaching contribution sometimes directly reflects the actual intraband EDF of pump-excited carriers (Fig. 3 and Fig. 5) [45, 48].

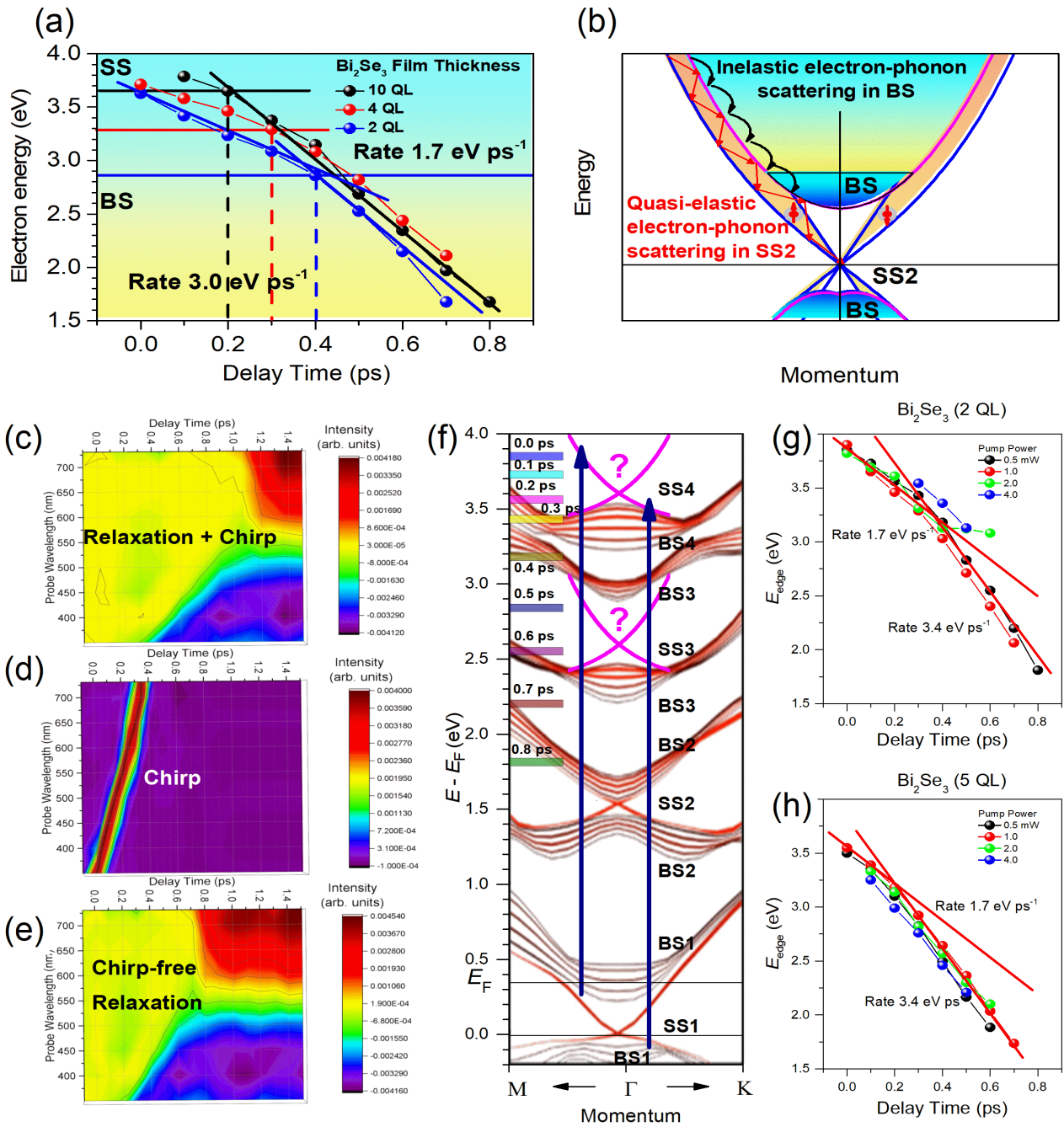

FIG. 5. (a) and (b) Electron energy distribution function (EDF) and the corresponding electron energy loss rates for several TI $Bi_2Se_3$ samples of different thickness, as indicated, measured with the two-photon pumping at ~ 730 nm (~ 1.7 eV) and the mechanisms of electron energy relaxation in Dirac surface states and bulk states, respectively. Reprinted with permission from [45]. Copyright (2021) American Chemical Society. (c), (d), and (e) Representative pseudocolor TA spectra plot of the 2D TI $Bi_2Se_3$ convoluted with the temporal chirp of the supercontinuum probing pulse, the temporal chirp itself and the resulting chirp-free TA spectra, respectively. The spectra were measured with the 340 nm (~ 3.65 eV photon energy) pumping. (f) Band structure of the 6 QL thick 3D TI $Bi_2Se_3$ film and the pumping transitions. The bulk and surface states are marked as BS and SS, respectively. The predicted in [46] higher energy Dirac SS are marked as "?". (g) and (h) Electron EDF for two samples of different thickness, as indicated. The linear fits (red color straight lines) and the corresponding electron energy loss rates are shown. The multicolor bars and the corresponding numbers in ps shown in (f) present the same data as in (g) for the 0.5 mW pumping power. Reproduced from [48]. CC BY 4.0.

Despite the diversity of models, in all of them the initial change in the absorption spectrum is due to two contributions associated with bandgap renormalization (the transient electric field model) or absorption bleaching (Drude-Lorentz model for excitons or the Fermi-Dirac statistics model for excitons and free carriers). A further decrease in the bandgap renormalization contribution reflects the weakening of the transient electric field. This behavior is due to a decrease in the carrier density caused by the formation of excitons or the recombination of electrons and holes. The decrease in the absorption bleaching contribution reflects the cooling of excitons or free carriers due to the cascade emission of photons, which is also followed by carrier recombination.

Since in most 2D semiconductors the exciton effect dominates over the free carrier effect, the absorption bleaching contribution is mainly determined by excitons and can be considered within the framework of the Drude-Lorentz model [58, 61]. However, since this model does not consider the quantum confinement of carriers and their Fermi-Dirac statistics (Pauli blocking), the corresponding polarization response cannot be used to characterize the dynamics of nonequilibrium excitons in 2D semiconductors. As for the dynamics of free carriers [9, 10], the Drude-Lorentz model is in principle inapplicable to characterize the transient absorption spectra. Instead, the Fermi-Dirac statistics model describes well the absorption bleaching contribution, regardless of whether it is caused by equilibrium or nonequilibrium carriers or excitons. The corresponding ultrafast carrier dynamics reflects Pauli blocking in the conduction band, as shown for TI $Bi_2Se_3$ [45–49]. Since these dynamics are extremely sensitive to the states occupied by the carriers, the absorption bleaching contribution follows the ultrafast carrier dynamics starting from their initial energy at pumping. Accordingly, this contribution shows a rapid growth with a rise-time constant of 0.1 – 0.3 ps, which directly characterizes the thermalization of the electron-hole plasma [13, 24, 44–49].

The decay of the absorption bleaching signals (<2–3 ps) reflects the cooling of photoexcited carriers or excitons due to their inelastic scattering on longitudinal optical (LO) phonons (Fröhlich relaxation mechanism) for polar semiconductors [13, 25, 28, 39, 45, 47-49, 67, 77-79] or non-polar optical phonons for non-polar semiconductors [80]. These cooling dynamics sometimes manifest themselves as a modulation of the overall decay trend in the pump-probe traces at the optical phonon frequency (coherent LO-phonons) [13, 39, 45, 47-49, 81]. In general, the carrier cooling rate does not depend on the density of photoexcited carriers [7, 8, 13, 82, 83]. However, the decay of LO-phonons occurs via an anharmonic three-phonon process involving acoustic phonon branches (the Clemens/Ridley process) [84]. Accordingly, the accumulation of acoustic phonons in 2D semiconductors with increasing density of photoexcited carriers effectively suppresses the Clemens/Ridley decay channel, thereby reducing the cooling rate of carriers. This process is known as the LO-phonon bottleneck and slows down the relaxation dynamics of carriers [28, 85]. For this reason, to achieve maximum carrier mobility in 2D semiconductors, a high sound speed is required [57] to avoid the LO-phonon bottleneck.

When the energy of carriers or excitons becomes less than the energy of optical phonons, further cooling occurs due to a less energy-consuming process associated with the emission of acoustic phonons [25, 44, 67, 77, 78, 80, 86, 87]. This behavior is in good agreement with the fact that the highest carrier mobility can be achieved for 2D semiconductors with high optical phonon energy [57] to avoid the Fröhlich carrier scattering mechanism when using high current densities. The typical relaxation time of photoexcited carriers due to the emission of acoustic phonons is ~10 ps [86, 87]. The acoustic phonon effect typically manifests itself as a modulation of the overall decay trend in the pump-probe traces at the acoustic phonon frequency (coherent acoustic phonons) [86, 87]. However, this stage of carrier energy relaxation is greatly overlapped by the relaxation dynamics associated with electron-hole recombination. The latter dynamics are typically bi-exponential due to nonradiative and radiative recombination, which occur within a few ps and a few hundred ps, respectively [8, 20, 88]. The existence of two electron-hole recombination mechanisms is determined by selection rules that follow from conservation of the momentum depending on whether the carriers relax to the band edges or not [62].

As we have already mentioned above, the influence of the transient electric field on the electronic structure of 2D semiconductors leads to bandgap renormalization [7, 8, 11, 50, 82, 89, 90]. In 3D semiconductors this effect is usually associated with a screening process, involving many-body effects, such as a correlated motion of carriers and their scattering with ionized impurities or optical phonons [7, 8, 11, 82]. As the 2D limit is approached, the influence of transient electric fields in depletion regions near the boundaries (quantum-confined Stark effect) [89] and the non-parabolicity of the conduction/valence band (quasiparticle mass renormalization) [90] can also contribute to the bandgap renormalization effect.

Despite the highly nonequilibrium population of photoexcited carriers, the changes in the absorption spectrum are linear in the strength of the transient electric field (linear electro-optic effect, Franz-Keldysh effect or the Stark effect). This behavior is in good agreement with the tuning of the bandgap of TMDCs by applying an electric field [32]. However, when using stronger pumping, the quadratic dynamics of the change in the transient absorption spectrum with respect to the transient electric field may occur (Kerr effect) [11, 62, 91, 92].

Since the positive broadband contribution due to bandgap renormalization is caused by the shift of the conduction band towards lower energies and, hence, the narrowing of the bandgap, it begins to manifest itself just below the bandgap energy [Fig. 1(d) and Fig. 4] [7, 8, 11, 82]. The rise-time of the corresponding pump-probe signal in this case is typically 0.5 – 1.0 ps [8], which is slightly longer than for the absorption bleaching signal. The longer growth trend indicates that the dynamics of pump-excited carriers, which causes the integral effect, are slower than the dynamics associated with the carrier population or exciton formation. The decay of the bandgap renormalization contribution (a few hundred ps) is due solely to the recombination of electrons and holes [88]. Accordingly, at longer delay times the relaxation tendency of this contribution is the same as for the absorption bleaching contribution.

**3. Physical phenomena determining contributions to the transient absorption spectra of 2D semiconductors.**

*3.1 Drude-type FCA and Drude-type FCA bleaching.*

As we mentioned above, one way to study the ultrafast dynamics of photoexcited carriers in 2D semiconductors is to measure the absorption by the carriers themselves. The corresponding contribution to the transient absorption spectra appears in the THz - mid-IR range and is due to Drude-type processes. Specifically, for pump-excited extrinsic carriers in doped 2D semiconductors (below-bandgap pumping), this contribution can be either positive or negative depending on whether Drude-type FCA or Drude-type FCA bleaching dominates (Fig. 2) [5]. In contrast, for pump-excited carriers in intrinsic 2D semiconductors (above-bandgap pumping), only Drude-type FCA contributes to the transient absorption spectrum, providing an exclusively positive response [6, 7]. Accordingly, near-IR - UV pumping and THz - mid-IR probing is commonly used to monitor the Drude-type FCA in intrinsic 2D semiconductors.

Despite the apparent simplicity of this absorption process, free (non-interacting) electrons in the conduction band cannot be optically excited to higher energy states of the band (intraband optical transitions) due to energy-momentum conservation restrictions [62]. However, this type of optical transition becomes allowed if they are accompanied by scattering carriers on the carriers themselves, as well as on phonons, defects or ionized impurities. The scattering process is necessary to conserve the electron wavevector during the optical transition and is therefore a key factor controlling the efficiency of the Drude-type FCA.

The corresponding rate of energy absorption through an area element $\Delta A$, perpendicular to the incident radiation with intensity $I_\nu$ and frequency $\nu$ along the distance traveled by photons $ds$ in time $dt$, can be expressed in two manners, namely, using the intensity of incident light or the difference between the intensities of incident and transmitted light [93]:

$$\frac{dE}{dt} = I_\nu \alpha_{D0} dA ds = -dI_\nu dA, \quad (1)$$

where $\alpha_{D0}$ is the Drude-type FCA coefficient, expressed in units of cm$^{-1}$ [62],

$$\alpha_{D0} = \frac{\varepsilon_\infty \nu_p^2 \gamma_c}{n_r c (\nu^2 + \gamma_s^2)}, \quad (2)$$

with

$$\nu_p = \left(\frac{1}{2\pi}\right)\sqrt{\frac{4\pi n_e e^2}{m_e^* \varepsilon_\infty}} \quad (3)$$

being the corresponding plasma frequency of free carriers screened by the high-frequency dielectric constant $\varepsilon_\infty$, and where $n_e$ is carrier density, $c$ is speed of light, $\gamma_s$ is the scattering rate of carriers due to various scattering mechanisms, $e$ is the electron charge, $m_e^*$ is the electron effective mass, and $n_r$ is the real part of the refractive index, which is approximately independent of $\nu$.

The magnitude of the Drude-type FCA coefficient varies with the frequency of the light, thereby completely determining the frequency range in which the probe light can be absorbed. Accordingly, the efficiency of the absorption process strongly depends on the electron scattering rate and the free carrier density. Typical scattering rates ($\gamma_c$) for carrier-carriers scattering (effective interaction time ~50 fs), carrier-optical-phonon scattering and carrier-ionized impurity scattering (effective interaction time ~1 ps) and carrier-acoustic-phonon scattering (effective interaction time ~10 ps) can be estimated as ~20, ~1.0, ~0.1 THz, respectively [8, 13, 45]. On the other hand, since in dopped semiconductors the density of extrinsic free carriers can be in the range $n_e$ ~$10^{16}$ – $10^{19}$ cm$^{-3}$, $\nu_p$ usually lies in the THz - mid-IR range [62]. Although the typical photoexcited carrier density in intrinsic 2D semiconductors is significantly higher ($n_e$~$10^{18}$ – $10^{20}$ cm$^{-3}$), $\nu_p$ still lies in the mid-IR range [8]. For example, for a photogenerated carrier density of ~$10^{20}$ cm$^{-3}$, it can be estimated as ~100 THz. Thus, the obtained value significantly exceeds the carrier scattering rates ($\gamma_s \ll \nu_p$).

Integrating Eq. (1) over the laser pulse duration yields the differential equation,

$$\frac{dI_\nu}{I_\nu} = -\alpha_{D0} ds, \quad (4)$$

which in turn leads to a solution identical to Beer's law,

$$I_\nu = I_{\nu 0} e^{-\alpha_{D0} d}, \quad (5)$$

where $d$ denotes the thickness of the 2D semiconductor. Accordingly, for photoexcited carriers in intrinsic 2D semiconductors (above-bandgap pumping), the change in optical density [OD $\equiv \ln(I_{\nu 0}/I_\nu)$] due to Drude-type FCA is as follows:

$$\Delta \text{OD}^{DA} = \alpha_{D0} d = \frac{c r_e n_e \gamma_s d}{\pi n_r (\nu^2 + \gamma_s^2)}, \quad (6)$$

where $r_e = e^2/(m_e^* c^2)$ is the classical electron radius. Thus, this contribution to the transient absorption spectrum is always positive.

In contrast, for photoexcited extrinsic carriers in doped 2D semiconductors (below-bandgap pumping), $\Delta \text{OD}^{DA}$ is read as

$$\Delta \text{OD}^{DA} = \Delta \alpha_D d, \quad (7)$$

where $\Delta \alpha_D = \alpha_D - \alpha_{D0}$ is the change in the Drude-type FCA coefficient caused by pumping. In this case, the pump-modified Drude-type FCA coefficient can be expressed as

$$\alpha_D = \alpha_{D0}(1 - f_e), \quad (8)$$

where

$$f_e = \frac{1}{\exp\left[\frac{h\nu_{probe} - E_F}{k_B T_e}\right] + 1} \quad (9)$$

is the Fermi–Dirac occupancy factor for electrons with $E_F = \frac{h^2}{8\pi^2 m_e}(3\pi^2 n_e)^{2/3}$ and $k_B$ being the Boltzmann constant [8]. Accordingly, the contribution associated with the Drude-type FCA to the transient absorption spectrum can be either positive [Eq. (6)] or negative,

$$\Delta \text{OD}^{DB} = -\alpha_{D0} f_e d, \quad (10)$$

depending on whether the conduction band states below $E_F$ are filled or not [Fig. 2(e) and (f)]. The negative contribution describes the Drude-type FCA bleaching. It is worth noting that in most cases, due to the spatial redistribution of photoexcited extrinsic carriers, both the Drude-type FCA [Eq. (6)] and the Drude-type FCA bleaching [Eq. (10)] act simultaneously [Fig. 2(a)].

It is important to note that while the amplitude of the Drude-type FCA contribution to the transient absorption spectrum depends linearly on $n_e$ [Eq. (6)], the dependence of the Drude-type FCA bleaching contribution is weaker, since the dependence on $n_e$ is more complex [Eqs. (9) and (10)] [8],

$$\Delta \text{OD}^{DB} \propto n_e^{1/6}. \quad (11)$$

To analyze the dynamics of carriers in the frequency domain, two limiting cases can be considered. For the low-frequency limit ($\nu \ll \gamma_s$), $\alpha_{D0}$ of Eq. (2) can be rewritten as

$$\alpha_{D0} = \frac{\varepsilon_\infty v_p^2}{n_r c \gamma_s} = \frac{c r_e n_e}{\pi n_r \gamma_s} \quad . \tag{12}$$

According to the above estimate $\gamma_s \ll v_p$, the Drude-type FCA in the low-frequency limit contributes to the transient absorption spectrum in the region below $v_p$. Thus, the intraband optical transitions in 2D semiconductors are allowed exclusively in the frequency range below $v_p$, and their efficiency is as high as $\gamma_s$ is low [Fig. 1(a)].

Equation (12) also describes the time domain behavior. For photoexcited extrinsic carriers in doped 2D semiconductors, the temporal dynamics of the Drude-type FCA are controlled exclusively by $\gamma_s$, since $n_e$ does not change during pumping. In contrast, for photoexcited carriers in intrinsic 2D semiconductors, the decay of the Drude-type FCA response is controlled by both quantities. Thus, the initial cooling of photoexcited carriers in both cases is completely determined by $\gamma_s$ associated with optical phonons. The reason for this behavior is that the scattering rate of carriers on optical phonons increases with decreasing carrier energy [62]. Accordingly, the cooling time of carriers is determined by the initial energy of the carriers and the number of acts of their electron-phonon scattering. Since the scattering time of an electron on an optical phonon is several tens of femtoseconds [13, 45], the cooling time of carriers is usually several ps. The influence of acoustic phonons on the initial relaxation of carriers is insignificant. For photoexcited carriers in intrinsic 2D semiconductors, the further decay of Drude-type FCA is controlled by the rate of electron-hole recombination, which significantly exceeds the rate of carrier scattering by acoustic phonons.

For the high-frequency limit ($v \gg v_p \gg \gamma_s$), the Drude-type FCA coefficient can be expressed as

$$\alpha_{D0} = \frac{\varepsilon_\infty v_p^2 \gamma_s}{n_r c v^2} = \frac{c r_e n_e \gamma_s}{\pi n_r v^2}. \tag{13}$$

In this case, the Drude-type FCA coefficient sharply tends to zero regardless of $\gamma_s$ and $n_e$. This behavior means that intraband optical transitions in 2D semiconductors are strictly forbidden in the near-IR - UV ranges (200 - 1000 THz) [Fig. 1(a)]. In other words, the photoexcited electron population becomes transparent in the high-frequency range, since in this case the free electrons are unable to absorb light due to energy-momentum conservation restrictions. However, there is one exception to the high-frequency limit, namely $v \gg \gamma_s \gg v_p$, which has never been considered due to the unrealistically high electron scattering rate. However, in this case the Drude-type FCA coefficient becomes highly dependent on the electron scattering rate and can potentially extend the FCA range towards higher frequencies.

To summarize this section, we note that the contributions to the transient absorption spectra of 2D semiconductors for pumping below/above the bandgap and probing below the bandgap should be considered within the framework of the Drude model. The frequency range of Drude-type FCA is limited by the plasma frequency and strongly depends on the free carrier density and the electron scattering rate. The most specific case occurs at the high frequency limit and the extremely high electron scattering rate. In this case, the high-frequency edge for Drude-type FCA can be extended towards higher frequencies. This behavior is unrealistic in 3D semiconductors, whereas in 2D intrinsic semiconductors it becomes possible, as discussed in the next section.

### 3.2 Inverse-bremsstrahlung-type FCA.

For intrinsic 2D semiconductors, when pumping and probing above the bandgap, there is an additional effect that causes absorption of light by photoexcited carriers and, therefore, leads to a positive contribution to the transient absorption spectrum [Fig. 1(c)] [45-49]. This effect arises due to the strong vertical confinement of photoexcited carriers with large momentum. The high kinetic energy and mobility of photoexcited carriers leads to their collisions with potential barriers at the boundaries of 2D semiconductors (Fig. 5). The effective interaction time of carriers due to collisions is limited to ~1 fs [50]. Accordingly, this scattering process is characterized by an exceptionally high carrier scattering rate (≤1000 THz), which significantly exceeds all the scattering rates of carriers considered in the previous section for Drude-type FCA. Despite the similarity with Drude-type FCA, the corresponding absorption process in intrinsic 2D semiconductors is called inverse-bremsstrahlung-type FCA [45]. From the point of view of the Drude model, collisions of carriers with potential barriers at the boundaries of intrinsic 2D semiconductors are much more probable than scattering of carriers on the carriers themselves, as well as on phonons, defects, or ionized impurities.

Inverse-bremsstrahlung-type FCA in intrinsic 2D semiconductors is considered as an analogue of what is well known in collisional plasma physics. Accordingly, the corresponding rate of energy absorption is determined by the rate of energy gain by an electron in a collisional plasma exposed to an intense laser field of amplitude $E_0$ [94]:

$$\frac{dE}{dt} = \frac{e^2 E_0^2 v_{\text{eff}}}{2m_e (v^2 + v_{\text{eff}}^2)} \quad , \tag{14}$$

where $v_{\text{eff}}$ denote the effective collision frequency, which increases with the carrier energy and is equivalent to the carrier scattering rate in the Drude model. Accordingly, the rate of absorption of laser radiation by a collisional plasma of density $n_e$ can be expressed as

$$\frac{dE}{dt} = \frac{4\pi c r_e I_v v_{\text{eff}} n_e}{(v^2 + v_{\text{eff}}^2)} dA ds \quad , \tag{15}$$

where $I_v = (c/8\pi) E_0^2$ is the laser light intensity.

Similarly, as in the case of Drude-type FCA, a comparison of the energy absorption rates yields the following:

$$\frac{dE}{dt} = \frac{4\pi c r_e I_v v_{\text{eff}} n_e}{(v^2 + v_{\text{eff}}^2)} dA ds = I_v \alpha_{IB} dA ds = -dI_v dA. \tag{16}$$

Integrating Eq. (16) over the laser pulse duration, the attenuation of light in a collisional plasma due to inverse-bremsstrahlung-type FCA can be expressed similarly to Eq. (5), but with an absorption coefficient of the following type:

$$\alpha_{IB} = \frac{4\pi c r_e n_e v_{\text{eff}}}{(v^2 + v_{\text{eff}}^2)} \quad . \tag{17}$$

For the high-frequency limit ($v \gg v_{\text{eff}}$), Eq. (17) simplifies to

$$\alpha_{IB} = \frac{4\pi c r_e n_e v_{\text{eff}}}{v^2} \quad , \tag{18}$$

which is basically the same as Eq. (13). Accordingly, the absorption caused by inverse-bremsstrahlung-type FCA drops sharply with increasing probe light frequency in the same way as for Drude-type FCA. However, the carrier scattering rate in the latter case is significantly lower than the effective collision frequency for inverse-bremsstrahlung-type FCA. This behavior means that the decrease in the efficiency of the inverse-bremsstrahlung-type FCA with increasing light frequency will not be as sharp as in the case of the Drude-type FCA, thereby extending the FCA towards a higher frequency range.

This agreement between the theoretical results obtained using two completely different approaches reflects the fundamental role of carrier collisions for the light absorption process in 2D semiconductors. As discussed in the previous section, due to the high electron scattering rate (effective collision frequency) in 2D semiconductors, any restrictions on simultaneous energy and momentum conservation for intraband optical transitions can be lifted over a much wider spectral range compared to Drude-type FCA.

Inverse-bremsstrahlung-type FCA can occur both during pumping and during probing. However, since the spectral intensity of the pump light is significantly higher than that of the probe light, this mechanism is more efficient in pumping. For this reason, the inverse-bremsstrahlung-type FCA is the basis for the realization of multiphoton-pumped UV-Vis transient absorption spectroscopy of 2D semiconductors [Fig. 1(c) and (d)] [50]. The multiphoton (multistep) pumping regime in this case means that, due to the extremely high scattering rate of carriers in 2D semiconductors, photoexcited carriers experience multiple collisions with potential barriers at the boundaries of 2D semiconductors during the pump pulse. This behavior results in the absorption of additional photons by photoexcited carriers and their acquisition of extremely high kinetic energy, which in turn increases the carrier scattering rate. Accordingly, the higher the energy of the pump photons, the higher the kinetic energy of the photoexcited carriers and the rate of their scattering.

Carrier collisions continue to occur after pumping is complete until the carriers lose their energy and reach the band edges, where they recombine. Consequently, for the probe beam, inverse-bremsstrahlung-type FCA occurs when the probe light interacts with the relaxing carriers during their collisions. The corresponding positive contribution to the transient absorption spectra is as follows:

$$\Delta \text{OD}^{IB} = \alpha_{IB} d \ . \qquad (19)$$

Since the amplitude of this contribution depends linearly on $n_e$ (and hence on the pump power) and $\nu_{\text{eff}}$ (and hence on the pump photon energy), the FCA range can be extended to near-IR or even Vis range by increasing both quantities (Fig. 3).

Thus, in intrinsic 2D semiconductors, the unusual carrier dynamics lead to the inverse-bremsstrahlung-type FCA, which is responsible for multiphoton (multistep) pumping. This effect extends the transient absorption spectra of intrinsic 2D semiconductors towards higher energies, significantly exceeding the energy of pump photons. In addition, the inverse-bremsstrahlung-type FCA can make a significant contribution to the transient absorption spectrum with an increase in the density of pump-excited carriers and their energy. The corresponding initial relaxation of carriers combines their multiple collisions with potential barriers at the boundaries of 2D semiconductors and scattering by optical phonons.

*3.3 Bandgap renormalization.*

As we mentioned above, the bandgap renormalization effect is a very complex phenomenon. However, it can be considered using a simple phenomenological approach that was originally proposed to characterize this effect in thin-film semiconductors [7]. Accordingly, the broadband positive contribution to the transient absorption spectra with an onset just below the bandgap energy is due to new unoccupied states that were created by shifting the entire conduction band toward lower energy (bandgap narrowing) [Fig. 1(d) and Fig. 4]. This behavior is expected to occur with negligible change in the effective electron mass. Then further dynamics with decreasing film thickness to the 2D limit can be associated with the renormalization of the effective mass of quasiparticles [90].

To convert the bandgap narrowing into the frequency domain, we consider the absorption coefficient without pumping in its traditional form [95]:

$$\alpha_0 = C \frac{(h\nu_{probe} - E_g)^{1/2}}{h\nu_{probe}} \ , \qquad (20)$$

where $C$ is a constant. Then, similarly to how it was done for the Drude-type FCA [Eq. (7)], the corresponding contribution to the transient absorption spectrum can be expressed as

$$\Delta \text{OD}^{BGR} = (\alpha_{BGR} - \alpha_0) d \ , \qquad (21)$$

which is positive, originates below the bandgap energy, reaches a maximum slightly above it, and gradually decreases towards higher energies [Fig. 1(d) and Fig. 4]. Since the magnitude of bandgap narrowing ($\Delta E^{BGR}$) is known to be proportional to the Fermi wavevector $k_F = \sqrt{E_F 2 m_e^*}/\hbar$ [8],

$$\Delta \text{OD}^{BGR} = \frac{Cd}{h\nu_{probe}} \left[ \left( h\nu_{probe} - E_g + \Delta E^{BGR} \right)^{1/2} - \left( h\nu_{probe} - E_g \right)^{1/2} \right] \approx \frac{Cd \Delta E^{BGR}}{h\nu_{probe} (h\nu_{probe} - E_g)^{1/2}} \propto k_F \qquad (22)$$

and hence

$$\Delta \text{OD}^{BGR} \propto n_e^{1/3} \ . \qquad (23)$$

The latter proportionality with respect to the carrier density is generally accepted for the bandgap renormalization effect [7, 8, 11, 82].

Since the high-energy tail of the broadband positive contribution associated with bandgap renormalization overlaps with the narrower negative absorption bleaching peaks, the positive contributions are energetically located below and above them. This behavior determines the multi-peak structure of the transient absorption spectrum (Fig. 4). Although Eq. (22) correctly describes the strongly asymmetric trend of the bandgap renormalization contribution, its application to fitting transient absorption spectra appears to be difficult. Accordingly, for these purposes, one can use the asymmetric Gaussian function [Fig. 4 (right)],

$$\Delta\text{OD}^{BGR} = \frac{A}{\sqrt{2\pi\sigma^2}} e^{-\frac{(h\nu_{probe}-E_g)^2}{2\sigma^2}}$$
$$\times \left[1 + \text{erf}\left(\beta \frac{h\nu_{probe}-E_g}{\sqrt{2\sigma^2}}\right)\right], \quad (24)$$

where $A$, $\sigma$ and $\beta$ are the fitting parameters. Despite the significant simplification of the absorption edge modeling in this case, excellent agreement with the experimental transient absorption spectra can be obtained.

Thus, the contribution of the bandgap renormalization to the transient absorption spectrum is the result of the integral influence of pump-excited carriers on the electronic structure of 2D semiconductors, which is realized through the corresponding transient electric field. This effect provides a broadband positive contribution that starts at an energy just below the bandgap, peaks just above it, and gradually decreases with increasing energy.

*3.4 Absorption bleaching in 2D intrinsic semiconductors (Pauli blocking of the conduction and valence band states).*

Finally, we consider the negative contribution to the transient absorption spectra of 2D intrinsic semiconductors. Observing this contribution requires pumping and probing with photon energies exceeding the bandgap energy of 2D semiconductors, or resonant pumping and probing of exciton states [7–10, 20–25, 27, 28, 30, 45–49, 51–53, 68]. This contribution is due to the Fermi-Dirac statistics of pump-excited carriers in 2D semiconductors and reflects the population dynamics of free electrons and holes in the quasi-continuous conduction and valence bands, respectively (Fig. 3 and Fig. 4). Although the behavior of bound electron-hole pairs (excitons) is described by Bose-Einstein statistics, the formation of both equilibrium and nonequilibrium excitons is determined by Fermi-Dirac statistics. Accordingly, if the existing electron and hole states are already occupied by other excitons, the formation of new excitons is suppressed due to Pauli blocking.

As we mentioned above, this behavior leads to absorption bleaching in 2D semiconductors in the spectral range just above the bandgap or at exciton resonance energies [Fig. 1(b) and (c) and Fig. 4]. In other words, the optical bandgap for the probe light is dynamically extended because the states at the band edges are occupied by pump-excited carriers [7, 8, 82]. Accordingly, optical transitions from empty quantum states in the quasi-continuous valence band or to occupied quantum states in the quasi-continuous conduction band are blocked due to the Pauli exclusion principle. The corresponding dynamics of carrier relaxation, as well as their accumulation at the edge of the conduction/valence band, is manifested in the redistribution of the intensity of this broadband contribution over time toward lower energies, thereby characterizing the intraband carrier EDF (Fig. 3 and Fig. 5) [45, 48]. Due to the high sensitivity of this contribution to the electron population dynamics, it reflects the internal structure of the valence or conduction band and its splitting caused by strong spin-orbit coupling and the absence of inversion symmetry [20–24] or the Rashba effect [8], respectively.

Just as was used for the Drude-type FCA [Eq. (1)], the energy absorption rate in intrinsic 2D semiconductors can be expressed as

$$\frac{dE}{dt} = I_\nu \alpha_0 dA ds = -dI_\nu dA, \quad (25)$$

where $\alpha_0$ is the corresponding absorption coefficient without pumping [Eq. (20)]. Similarly, as in the previous sections, one can obtain the well-known Beer's law:

$$I_\nu = I_{\nu 0} e^{-\alpha_0 d}. \quad (26)$$

Accordingly, the contribution to the transient absorption spectrum is as follows:

$$\Delta\text{OD}^{AB} = \Delta\alpha d \quad (27)$$

where $\Delta\alpha = \alpha - \alpha_0$ denotes the pump-induced change in absorption coefficient. Accordingly, the absorption coefficient upon pumping is expressed as

$$\alpha = \alpha_0(1 - f_e - f_h), \quad (28)$$

where

$$f_e = \frac{1}{\exp\left[\frac{h\nu_{probe}-E_g-E_F}{k_B T_e}\right]+1} \quad (29)$$

and

$$f_h = \frac{1}{\exp\left[\frac{h\nu_{probe}+E_F}{k_B T_h}\right]+1} \quad (30)$$

are the Fermi–Dirac occupancy factors for electrons and holes, respectively, with $E_F$ measured from the top of the valence band.

The corresponding contribution to the transient absorption spectrum due to absorption bleaching can be expressed as follows [8]:

$$\Delta\text{OD}^{AB} = -\alpha_0(f_e + f_h)d, \quad (31)$$

which is negative once the pump-excited electrons and holes fill up the conduction and valence band states, respectively, in accordance with their occupancy factors. However, because usually $m_h \gg m_e$, the effect mainly reflects the electron dynamics. The amplitude of the corresponding contribution to the transient absorption spectra changes with $n_e$ similarly to Eq. (11), that is, approximately as [8]

$$\Delta\text{OD}^{AB} \propto n_e^{1/6}. \quad (32)$$

To summarize this section, we note that the negative contribution to the transient absorption spectrum of 2D semiconductors, measured using pumping and probing above the bandgap or resonant pumping and probing of exciton states, should be considered within the framework of the Fermi-Dirac statistics model. This contribution reflects the dynamics of the population of pump-excited free electrons and holes in the conduction and valence bands, respectively, or the formation of excitons. The corresponding optical response is due to absorption bleaching, which is caused by the Pauli blocking of conduction and valence band states or exciton states by pump-excited carriers.

**4. The most common errors that arise in measuring, processing, and interpreting transient absorption spectra of 2D semiconductors.**

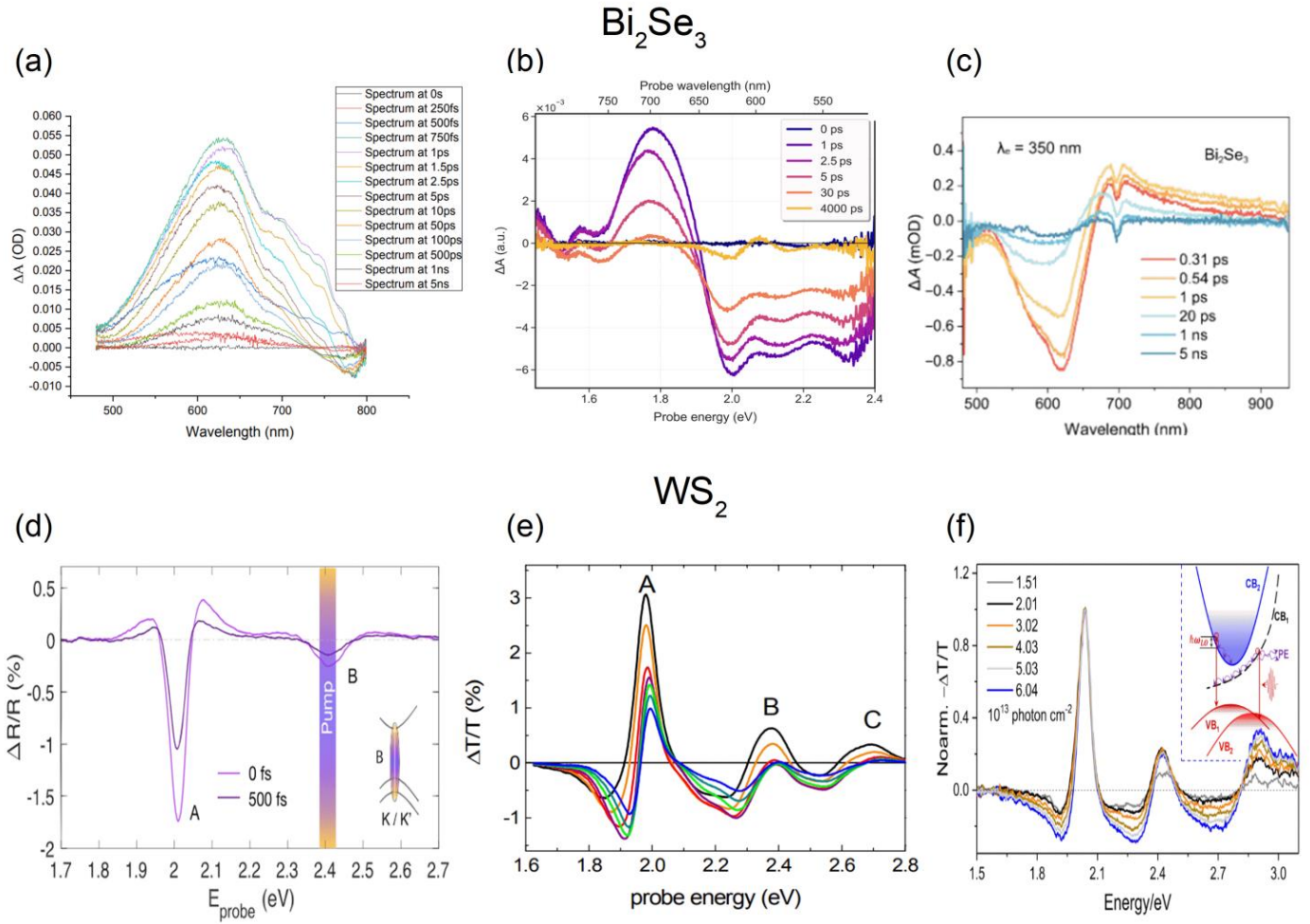

FIG. 6. (a), (b), and (c) Transient absorption spectra of 2D semiconductor $Bi_2Se_3$. Reprinted with permission from [51]. Copyright (2025) American Chemical Society; Reproduced from [52]. CC BY 4.0; Reproduced with permission from [53]. CC BY-NC 4.0. (d), (e), and (f) Transient absorption spectra of 2D semiconductor $WS_2$. (d) Reproduced from [25]. CC BY 4.0; (e) Reproduced from [30] with permission from the Royal Society of Chemistry; (f) Reproduced from [28]. CC BY 4.0.

As we mentioned in the introduction, the use of two completely different approaches to interpreting ultrafast transient absorption spectra leads to many confusions. Here we consider some of them, as well as those that arise when obtaining and processing experimental data. As an example, Figure 6 shows typical transient absorption spectra of TI $Bi_2Se_3$ and TMDC $WS_2$. Although the spectra in both cases characterize the same materials, their shape and the signs of the contributions may differ depending on the experimental methods and data processing used. Below, several reasons that may lead to such discrepancies in the transient absorption spectra and their interpretations are discussed.

*4.1 Wavelength to Energy scale conversion.*

Since the transient absorption spectrum typically covers a wide spectral range, to correctly convert data recorded in wavelength units (nm) to energy units (eV), it is necessary not only to convert the wavelength scale, but also to multiply the signal values by $\lambda^2$ [45-48, 96]. This procedure simply converts units of measurement in spectroscopy and nothing more. In other words, the shape of the transient absorption spectra changes significantly when converting units of wavelength to units of energy. Accordingly, the presentation of spectra in two scales simultaneously is methodologically incorrect, since the spectra presented in each of them look different. Transient absorption spectra not transformed in this way do not reflect the actual intensity distribution in the spectrum and do not allow comparison of spectra measured by different research groups.

*4.2 Sign of contributions in transient absorption spectra.*

As we mentioned in the introduction, the actual sign of the transient absorption spectrum can be easily changed to opposite by adjusting the chopper modulation phase. Accordingly, the presented transient absorption spectra can be either positive or negative, regardless of the experimental configuration used. Figure 6(a), (b), and (c) [51-53] and Figure 3 [48, 49] show transient absorption spectra of TI $Bi_2Se_3$. Despite the same material, the spectra look very different. This can be due to several reasons. First, we note that the optimal geometry for measuring transient absorption spectra is normal incidence of the supercontinuum probe beam [50]. Therefore, simultaneous measurements of transient transmission and transient reflectivity of 2D semiconductors seem to be not very accurate, since the shape of the spectra and their intensity depend on the angle of incidence of the probing beam.

Secondly, the mentioned diversity of transient absorption spectra indicates differences in their classification. In general, a steady-state transmission response is usually considered negative, while a steady-state absorbance response is considered positive, indicating a decrease in transmission (or an increase in absorbance) of incident light passing through the sample. The corresponding absorption coefficient $\alpha(\omega)$, where $\omega = 2\pi\nu$ denotes the angular frequency, is related to the imaginary part of the complex refractive index. Alternatively, the steady-state reflectivity response for photon energies above the bandgap is related to the real part of the complex refractive index [97]:

$$R(\omega) = \left[\frac{1-n(\omega)}{1+n(\omega)}\right]^2, \quad (33)$$

and is also positive. Thus, it is assumed that the steady-state absorption and steady-state reflection of semiconductors have the same sign [62, 98], since the Kramers-Kronig relations do not change the sign of the optical response [69]. This behavior also remains true for the corresponding changes $\Delta\alpha(\omega)$ and [97]

$$\Delta R(\omega) = \frac{4[n^2(\omega)-1]}{[n(\omega)+1]^4} \Delta n(\omega) \quad (34)$$

since

$$\Delta n(\omega) = 1 + \frac{c}{\pi} P \int_0^\infty \frac{\Delta\alpha(\omega')}{\omega'^2 - \omega^2} d\omega' \quad (35)$$

and

$$\Delta\alpha(\omega) = -\frac{4\omega^2}{\pi c} P \int_0^\infty \frac{\Delta n(\omega')-1}{\omega'^2 - \omega^2} d\omega', \quad (36)$$

where $P$ refers to the principal part of the integral.

These contribution signs remain valid for transient absorption spectra, but only for pump-induced absorption processes such as Drude-type FCA, inverse-bremsstrahlung-type FCA, and bandgap renormalization. In contrast, for the pump-induced transmission process (absorption bleaching), the sign of the contributions is opposite to that for the steady-state responses, namely, negative for transient absorption and reflection and positive for transient transmission. Readers can judge for themselves which contributions to the transient absorption spectra shown in Fig. 6 obey this sign rule and which do not.

It is important to note that despite the differences in the transient absorption spectra measured for TI $Bi_2Se_3$ [Fig. 6(a), (b), and (c) and Fig. 3], two contributions can be distinguished for all of them, associated with the filling of states in the conduction band by pump-excited carriers (absorption bleaching) and the absorption of light by pump-excited carriers (inverse-bremsstrahlung-type FCA).

*4.3 Application of the Drude-Lorentz model to characterize transient absorption spectra of 2D semiconductors.*

Theoretical modeling of the absorption edge of semiconductors is usually carried out by calculating their complex dielectric function within the framework of the Drude-Lorentz model (Lorentz oscillator model) [76]. Although this model is applicable near material resonances and treats carriers in a classical way, by assuming that they act independently, it is often used to model broadband transient absorption spectra of 2D semiconductors [58, 61] and thin-film semiconductors [9, 10]. It is worth noting that when using the Drude-Lorentz model for describing steady-state absorption spectra of bulk semiconductors, parameterization of the optical function is usually applied to achieve better agreement with experiment [99–101]. However, in many cases even these improved models provide acceptable results only over a very limited spectral range. Thus, the Drude-Lorentz model is very limited for modeling even steady-state absorption spectra of bulk semiconductors over a wide spectral range [62, 76].

The situation is even more complex for the transient absorption spectra of 2D semiconductors. If the transient absorption spectra due to bandgap renormalization or absorption bleaching are caused by nonequilibrium free carrier dynamics, then the "Lorentz oscillator model" cannot be used in principle. The reason for this is that in this case there are no oscillators that could be resonantly excited by the oscillating electric field of the supercontinuum probing beam. In addition, the pump-excited electron-hole plasma becomes very transparent for frequencies above the plasma frequency, which for typical excited carrier densities lies in the mid-IR range. In other words, free carriers stop responding to the oscillating electric field of the supercontinuum probe beam in the near IR – UV spectral range. The only exception when the Drude-Lorentz model can be applied to characterize the transient absorption spectra of 2D semiconductors is when the absorption bleaching is caused by exciton resonances. Moreover, within the framework of the Drude-Lorentz model, only the dynamics of equilibrium excitons can be considered. Instead, the mentioned Fermi-Dirac statistics model is applicable to describe the influence of both equilibrium and nonequilibrium excitons.

*4.4 The multi-peak structure of transient absorption spectra.*

The overlap of the contributions to the transient absorption spectra of 2D semiconductors discussed in the previous sections appears to be the source of confusion in their interpretation. In particular, the appearance of multiple positive and negative peaks (Fig. 4) [8, 20–25, 27, 28, 30] is the result of the superposition of contributions of opposite sign associated with absorption bleaching and bandgap renormalization. Specifically, the multi-peak structure of the transient absorption spectra of intrinsic 2D semiconductors in the near-IR - UV range is due to the superposition of a broadband contribution associated with bandgap renormalization and much narrower doublet contributions associated with absorption bleaching. The doublet structure arises from splitting of the valence or conduction band caused by strong spin-orbit coupling and lack of inversion symmetry [20–24] or the Rashba effect [8], respectively. Moreover, additional peaks may be caused by different onset positions of the contribution and different rise-/decay-time constants.

Despite the different bandgap energies for 2D semiconductors and the different nature of the splitting of the valence or conduction band, the transient absorption spectra exhibit a similar structure (Fig. 4). Negative features are usually associated with the conduction band absorption bleaching or the exciton absorption bleaching, whereas many different interpretations have been proposed for positive contributions [7–10, 20–25, 27, 28, 30, 51–53, 68]. Figure 4 (right) shows an example of fitting transient absorption spectra using asymmetric and normal Gaussian profiles for the bandgap renormalization and absorption bleaching effects, respectively. Accordingly, the multi-peak structure arises from the superposition of the broadband contribution associated with bandgap renormalization and the valence or conduction band splitting components associated with absorption bleaching. The lowest energy peak arises from the slightly different onset positions of these two contributions and the difference in their rise-time constants.

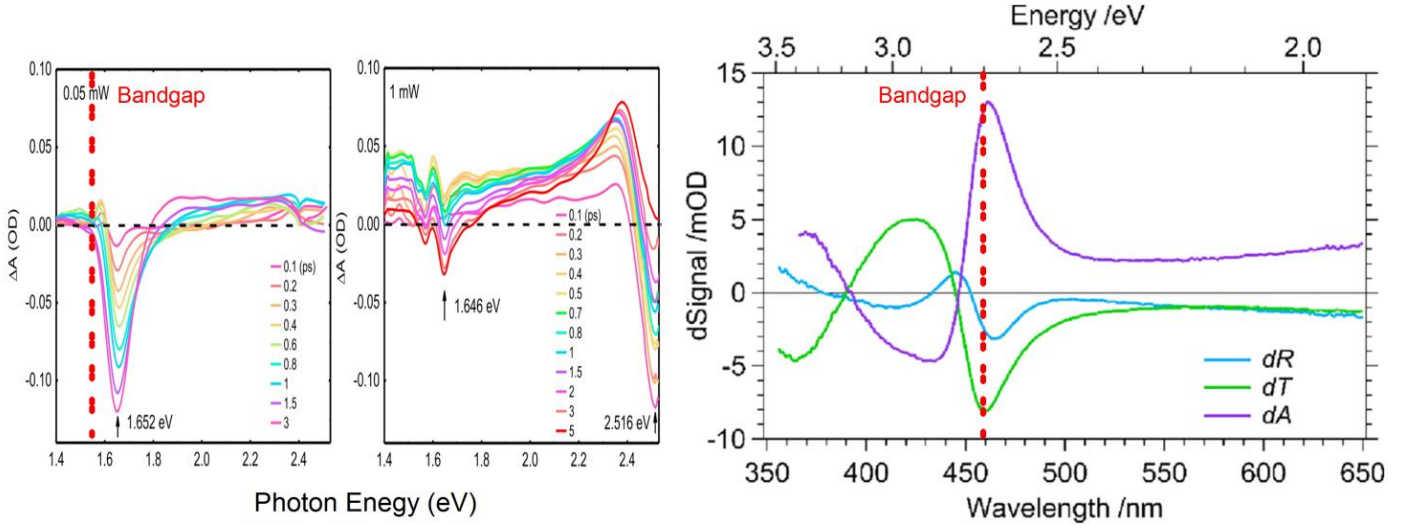

FIG. 7. (left) Transient absorption spectra of MAPbI$_3$ under pump power of 0.05 mW and 1 mW, correspond to carrier concentration of $4.66 \times 10^{17}$ cm$^{-3}$ and $9.32 \times 10^{18}$ cm$^{-3}$, respectively. Reproduced from [68]. CC BY 4.0. (right) Transient differential reflection (dR), transmission (dT), and absorption (dA) of a monoclinic BiVO$_4$ thin film (50 nm on fused silica) probed at 430 fs pump-probe delay and the incidence angle of 20$^0$. Excitation wavelength: 350 nm, power: 600 nJ/ pulse, approximate beam diameter: 0.3 mm. Reprinted with permission from [9]. Copyright (2018) American Chemical Society. For both cases, the approximate position of the bandgap is shown.

*4.5 Subpicosecond carrier relaxation dynamics in 2D semiconductors and chirp correction.*

The dynamics of ultrafast relaxation of carriers in the subpicosecond time range is of particular interest for the development of new optoelectronic devices based on 2D semiconductors. This time scale reflects the relaxation of carriers due to the emission of a cascade of optical phonons (Fig. 5) [13, 25, 28, 39, 45, 47–49, 67, 77-79] and characterizes the intraband EDF of carriers. As we already mentioned in the introduction, to obtain a realistic relaxation trend, a post-experimental procedure is necessary to eliminate the chirp of the supercontinuum probe pulse (Fig. 5) [3, 48]. Typically, this involves numerically extracting any shift in zero time with the probe pulse wavelength. Accordingly, the intraband carrier EDF may be lost if the chirp extraction procedure is not applied correctly. Moreover, as shown for 2D TI Bi$_2$Se$_3$ [48], detailed carrier relaxation trends associated with optical phonons in surface and bulk states, as well as their coherent coupling, can also be lost. However, these relaxation dynamics are very important for characterizing the switching speed in optoelectronic devices. Moreover, the dynamics of optical phonons and their coherent coupling may be of crucial importance in the study of 2D superconductors, where this effect plays a decisive role in the physical processes associated with superconductivity [102].

*4.6 Comparison of the "solid-state" and "molecular" approaches to the interpretation of transient absorption spectra.*

As we mentioned above, the combination of the "solid-state" and "molecular" approaches often creates confusion when interpreting transient absorption spectra of 2D semiconductors. To illustrate this, we consider below two examples of transient absorption spectra of thin-film semiconductors, originally interpreted using the "molecular" approach.

The first example concerns the "excited state absorption" observed for a 250 nm thick MAPbI$_3$ perovskite film [Fig. 7 (left)] [68]. As discussed above, in the "solid-state" approach, such intraband transitions in the conduction band are strictly forbidden in the near-IR - UV spectral range because electron momentum is not conserved during optical transitions. This conclusion is valid even for thin-film semiconductors with "several conduction bands", similar to how it happens for quantum subbands in semiconductor QWs and 2D semiconductors. As we mentioned in the introduction, the quasi-continuous nature of the conduction/valence band results in a joint density of states rather than discrete sets of electronic states. Accordingly, intraband transitions between two conduction bands (subbands) within a quasi-continuous conduction band, even in the 2D limit, obey the same selection rules that are associated with the conservation of the in-plane carrier momentum (Fig. 1).

On the other hand, the experimental results presented in Figure 7(left) can be easily explained using the superposition of the two contributions to the transient absorption spectra discussed above, namely bandgap renormalization and absorption bleaching. Despite the slightly different spectral range, the discussed transient absorption spectra are very similar to the spectra measured for a single layer of 20 nm MAPbBr$_3$ nanocrystals [Fig. 4 (right)] [8]. Accordingly, all changes in the spectrum mentioned in [68] with increasing pump power are due to the superposition of the two mentioned contributions. Specifically, the negative contribution due to absorption bleaching, reaching a maximum at ~1.6 eV (slightly above the bandgap energy), was almost completely compensated by the corresponding positive contribution from bandgap renormalization with increasing pump power [Fig. 7 (left)]. This behavior is because the contribution of bandgap renormalization depends more strongly on the carrier density than the contribution of the absorption bleaching ($n_e^{1/3}$ compared to $n_e^{1/6}$). Since the positive contribution is much wider than the negative one, the peak of the bandgap renormalization effect appears at ~2.35 eV with increasing pump power.

In addition, the negative contribution, reaching a maximum at ~2.5 eV, is due to the absorption bleaching in the second conduction band [68]. The rapid decay of this contribution within 5 ps indicates a tendency for carriers to relax toward the lower conduction band states [8]. Since carriers are initially excited at ~2.8 eV, they first populate the second conduction band at ~2.5 eV and then accumulate in the lower conduction band at ~1.6 eV before recombination. Thus, the dynamics of the transient absorption spectra, presented in Figure 7(left), characterizes the intraband electron EDF, as we discussed in the previous sections.

The second example concerns the transient absorption spectra of a thin film of monoclinic $BiVO_4$ with a thickness of 50 nm [Fig. 7 (right)] [9]. First, we note that the spectra were measured at a probe beam incidence angle of 20° and therefore reflect additional conditions that the authors do not discuss. According to the authors, the positive contribution to the transient absorption spectrum just below the bandgap is due to "charge carrier absorption (more specifically by holes)", while the negative contribution just above the bandgap is due to "ground-state bleaching".

As we already mentioned for the previous example, according to the "solid-state" approach, the intraband transitions in the conduction/valence band are strictly forbidden in the near-IR - UV range of the spectrum. Thus, the strong absorption by pump-excited holes in the visible range is not consistent with this approach, whereas it can be explained by bandgap renormalization [7, 8, 11]. Moreover, it is surprising that the authors did not observe a contribution to the transient absorption spectra associated with pump-excited electrons. As for the "ground-state bleaching", in the solid-state approach, absorption bleaching can occur due to the filling of states of either the conduction band or the valence band. Accordingly, the transient absorption (transmission) spectra shown in Figure 7 (right) can be explained by the two contributions discussed in the previous sections, namely bandgap renormalization and absorption bleaching.

Thus, the application of the "molecular" approach to interpreting transient absorption spectra of thin-film semiconductors or 2D semiconductors introduces some confusion not only in terminology, but also in physical processes that determine the ultrafast dynamics of photoexcited carriers.

## 5. Conclusions and perspectives.

This review discusses recent experimental results on ultrafast transient absorption spectroscopy of thin-film semiconductors in a wide spectral range from a few THz to UV (~1000 THz). We do not intend to consider here the complete set of transient absorption spectra of 2D semiconductors and the experimental configurations in which they were measured. Instead, we discuss the most general features of these spectra based on a "solid-state" approach to their interpretation. Specifically, we considered all the physical phenomena characterizing the ultrafast relaxation dynamics of carriers in these extremely thin solid-state systems.

Compared with the "solid-state" theoretical approach originally proposed to characterize the influence of injected carriers on the optical properties of bulk semiconductors [11], the "solid-state" approach discussed here includes an additional effect, namely inverse-bremsstrahlung-type FCA. This effect is a specific feature of 2D semiconductors and is caused by collisions of photoexcited carriers with potential barriers at their boundaries. Inverse-bremsstrahlung-type FCA extends the range of FCA usually associated with Drude-type FCA to near-IR or even Vis range [45–50]. The spectral position of this contribution can be tuned by changing the initial energy of photoexcited carriers (pump photon energy), as well as the density of photoexcited carriers (pump power) and the thickness of the 2D semiconductors. In this regard, this effect requires further study for various 2D semiconductors, since it can find wide application in optoelectronic devices as modulators of light intensity at specific wavelengths. Furthermore, the ability to tune the FCA over a wide spectral range is also attractive for devices operating in the IR range, such as thermal detectors and emitters.

In addition, we highlight that inverse-bremsstrahlung-type FCA is responsible for multiphoton (multistep) pumping in ultrafast transient absorption spectroscopy of 2D semiconductors [50]. This excitation mechanism extends the transient absorption spectra towards higher energies, significantly exceeding the energy of pump photons [8, 45–49]. We emphasize here that this type of multiphoton (multistep) pumping contrasts with the Auger-type resonant pumping mechanism proposed for explaining the excitation of spatially localized PL and its upconversion in TMDC monolayers [31, 33-35]. Although both mechanisms lead to the upconversion of carrier excitation, inverse-bremsstrahlung-type FCA does not require the carrier 3D spatial confinement, as the Auger-type mechanism does in the case of quantum dots (0D semiconductors) [70, 72]. Thus, the multiphoton (multistep) carrier excitation based on inverse-bremsstrahlung-type FCA appears to be the most realistic mechanism for excitation of carriers in various 2D semiconductors, where typically only vertical quantum confinement occurs.

Another important result that we highlight here is that the multi-peak structure of the transient absorption spectra of intrinsic 2D semiconductors in the near-IR – UV spectral range is due to the superposition of contributions associated with bandgap renormalization and absorption bleaching. Specifically, the splitting of the valence/conduction band, as well as differences in the onset positions and rise-/decay-time constants of these contributions, lead to the appearance of a multi-peak structure due to the dynamic compensation of contributions.

Comparing the "solid-state" approach to interpreting the transient absorption spectra of 2D semiconductors with the "molecular" approach, we concluded that the latter introduces some confusion not only into the terminology, but also into the physical processes that characterize the ultrafast dynamics of photoexcited carriers. We also emphasized that the Drude-Lorentz model is applicable to characterize the ultrafast transient absorption spectra of 2D semiconductors only if they reflect the dynamics of equilibrium excitons. Instead, the Fermi-Dirac statistics model is applicable to describe spectra characterizing the dynamics of free carriers and both equilibrium and nonequilibrium excitons.

We also noted other common errors that arise when measuring and processing transient absorption spectra of 2D semiconductors. In particular, the optimal geometry for the probe beam is normal incidence, so the transient reflectivity spectra are not as accurate as the transient transmission (absorption) spectra. In addition, the shape of the transient absorption spectra changes significantly when converting wavelength units to energy units. We also emphasized that proper extraction of the chirp of the supercontinuum probe beam is a key factor in obtaining the transient absorption spectra characterizing the intraband electron EDF. The combination of these results with those obtained using TrARPES creates a powerful platform for the comprehensive study of ultrafast carrier dynamics in 2D semiconductors and electronic devices based on them.

The method of ultrafast transient absorption spectroscopy can also be applied to study the dynamics of coherent optical phonons and their coupling in 2D superconductors. Understanding and controlling this phenomenon could have a decisive impact on improving the superconducting properties of 2D materials and their interfaces.

**Data availability statement**

No new data were created or analyzed in this study.

**ORCID ID**

Yuri D Glinka  https://orcid.org/0000-0002-2267-0473